\documentclass[twocolumn,showpacs,preprintnumbers,amsmath,amssymb]{revtex4}
\usepackage{graphicx}
\usepackage{dcolumn}
\usepackage{bm}
\usepackage{amssymb}
\usepackage{epsfig}
\usepackage{hyperref}
\usepackage{float}
\newcommand{\be}{\begin{equation}}
\newcommand{\ee}{\end{equation}}
\newcommand\beq{\begin{eqnarray}}
\newcommand\eeq{\end{eqnarray}}

\begin{document}
\title{Axion Cosmology with Early Matter Domination}
\author{Ann E. Nelson$^{1}$}
\email{anelson@phys.washington.edu}

\author{Huangyu Xiao$^{1}$}
\email{huangyu@uw.edu}
\affiliation{$^1$Department of Physics, University of Washington,  Seattle, WA 98195-1560,USA}
\begin{abstract}
The default assumption of early universe cosmology is that the postinflationary  universe was  radiation dominated until it was about 47000 years old. Direct evidence for the radiation dominated epoch  extends back until nucleosynthesis, which began during the first second. However there are theoretical reasons to prefer a period of earlier matter domination, prior to nucleosynthesis, e.g. due to late decaying massive particles needed to explain baryogenesis. Axion cosmology is quantitatively affected by an early period of matter domination, with a different axion mass range preferred and greater inhomogeneity produced on small scales.
In this work we show  that such increased inhomogeneity can lead to  the formation of axion miniclusters in axion parameter ranges that are different from those usually assumed.    If the reheating temperature is below $58$ MeV, axion miniclusters can form even if  the axion field is present during inflation and has been previously homogenized. The upper bound on the typical initial axion minicluster mass is  raised from $10^{-10} M_{\odot}$ to $10^{-7} M_{\odot}$, where $M_{\odot}$ is a solar mass. These results may have consequences for indirect detection of axion miniclusters, and could  conceivably probe the thermal history of the universe before nucleosynthesis.
 \end{abstract}
\date{\today}
\maketitle

\section{Introduction}
The QCD axion, which was invented to solve the strong CP problem\cite{Peccei:1977hh,Peccei:2006as}, is a well-motivated candidate for dark matter. The axion mass and couplings are determined by a single parameter, the axion decay constant    $f_a$. Laboratory, astrophysical and cosmological bounds on     $f_a$ place it  well above the weak scale. As the axion mass and couplings are inversely proportional to $f_a$,  the axion must be extremely light, long lived, and weakly coupled.          

  If the axion exists, the misalignment mechanism  produces axion dark matter, with an abundance that  increases with $f_a$.  It is often stated that there is an upper bound on $f_a$ of $10^{12}$ GeV so as  to not overproduce axions. This bound may be relaxed, e.g., if the axion exists during inflation and our patch of the universe happens to have a small misalignment, or with a new depletion mechanism\cite{Agrawal:2017eqm}.   Without such tuning or depletion, the  allowed value of $f_a$ is in the window $10^9\text{GeV}<f_a<10^{12}\text{GeV}$\cite{Andriamonje:2007ew,Ayala:2014pea,Raffelt:2006cw,Preskill:1982cy,Abbott:1982af,Dine:1982ah,Vysotsky:1978dc,PhysRevD.80.035024}. It has been argued that string theory favors  a higher value of $f_a$ \cite{Banks:2003sx, Svrcek:2006yi, Conlon:2006tq} and lighter axion than this window allows.
 
 We can detect axions directly through the couplings with SM particles, especially the axion-photon coupling (For some reviews, see \cite{Graham:2015ouw,Bradley:2003kg}). However, there are  other interesting strategies for axion indirect detection.
 The axion can form gravitational bound states on small scales at very early times. If the axion is produced after inflation, then the axion field has an alignment angle which varies over a scale on the order of the Hubble horizon size of the universe at the time of formation\cite{Kibble:1976sj}. Such   inhomogeneities  can grow and become gravitational bound states called axion miniclusters \cite{Tkachev:1986tr,Hogan:1988mp,Sakharov:1994id,Khlopov:1998uj}. Axion miniclusters could  grow to bigger structures or boson stars\cite{Kolb:1993zz,Seidel:1994zb}, which could be detected by gravitational microlensing\cite{Fairbairn:2017dmf,Fairbairn:2017sil}.   On the other hand, if the axion exists  during inflation it is much more homogenous initially\cite{Kolb:1993hw,Hogan:1988mp,Kolb:1995bu,Chang:1998tb,Hardy:2016mns}. For some references on possible consequences and observations connected with axion miniclusters and axion stars see refs.~\cite{Barranco:2012ur,Berezinsky:2014wya,Tkachev:2014dpa,Tinyakov:2015cgg,Marsh:2015xka,Braaten:2016dlp,Levkov:2016rkk,Bai:2016wpg,Davidson:2016uok,Visinelli:2017ooc,Enander:2017ogx,Bai:2017feq,Iwazaki:2017rtb,Eby:2017xaw,Hertzberg:2018zte}, and for work on their structure and stability see refs~\cite{Seidel:1994zb,Barranco:2010ib,Braaten:2015eeu,Mukaida:2016hwd,Chavanis:2016dab,Eby:2016cnq,Helfer:2016ljl,JacksonKimball:2017qgk,Bai:2017feq,Michel:2018nzt}. For work on the possible unique signatures of axion structure formation due to their quantum mechanical properties as light degenerate bosons see refs.~\cite{Nambu:1989kh,Sikivie:2009qn,RindlerDaller:2009er,RindlerDaller:2011kx,RindlerDaller:2012vj,Saikawa:2012uk,Noumi:2013zga,Davidson:2013aba,Davidson:2014hfa,Guth:2014hsa}.

The properties of axion miniclusters sensitively depend  on the thermal history at the critical time when the axion starts to oscillate.  For a radiation dominated universe, the corresponding temperature is typically  about  1---10  GeV. This critical time is before big bang nucleosynthesis (BBN) and before the time when the big bang neutrinos decouple, and is during a time which is not connected to any established cosmological observable.  If we consider a different thermal history for the universe prior to a temperature of a few MeV, we will see that the upper bound on $f_a$ is relaxed, and  there is a significant difference in the formation history of axion miniclusters. With early matter domination, axion miniclusters can form even if the axion field has been homogenized by inflation, due to the more rapid growth of small scale primordial perturbations of the axion.  Such early growth of    substructure during early matter domination has been considered for other candidate dark matter particles\cite{Erickcek:2011us}. The  axion is special among dark matter candidates because its free streaming effects are almost negligible, so very small structures can form and survive.

In this paper we will consider the early cosmology of the standard invisible QCD axion with a nonstandard thermal history, with a period of early  matter domination prior to nucleosynthesis. Such matter domination can be due to a heavy, weakly coupled particle whose decays reheat the universe, as is required in some theories of low scale baryogenesis.
 We will briefly review the theory of the  axion and its corresponding cosmology, including the axion relic density and the formation of axion miniclusters in section \ref{s1}. In section \ref{s2} we will show how the axion window is opened  by early matter domination. In section \ref{s3}, a different story of axion minicluster formation with early matter domination is discussed. We will find that early matter domination potentially gives a larger initial characteristic mass of axion miniclusters.   

\section{Axion Cosmology}\label{s1}
Here we review the   axion and its cosmology. (For more details about axion cosmology, see\cite{Sikivie:2006ni,Marsh:2015xka,Arvanitaki:2009fg,diCortona:2015ldu,Visinelli:2017imh}.) The axion is a pseudo Nambu-Goldstone Boson  resulting from the spontaneous breaking of an approximate   symmetry  known as the Peccei-Quinn (PQ) symmetry, due to the vacuum expectation value of a complex field known as the PQ field.
We consider the following Lagrangian for the PQ field, which we call $\phi$:
\begin{equation}
\mathcal{L}_{\phi}=\frac{1}{2}\partial_{\mu}{\phi}^{\dagger}\partial^{\mu}\phi-\frac{\lambda}{4}(\phi^{\dagger}\phi-f_a^2)^2+....
\end{equation}
where the dots represent possible interaction terms with other particles and $f_a$ represents the vacuum expectation value of  $\phi(x)$. The   symmetry breaking will occur at a temperature $T_{\text{PQ}}$ which is roughly at the scale $f_a$. Classically, because of the PQ symmetry, the phase of $\phi$ is undetermined by the potential.   After the PQ symmetry breaking, the phase of the PQ field receives a small potential from nonperturbative QCD effects which is minimized at a value for which the strong CP violation vanishes. Fluctuations of the phase about the minimum are parameterized by the axion field $a(x)$.  Ignoring the energetically costly fluctations of the radial direction of $\phi$,   we may write
\begin{equation}
\langle\phi(x)\rangle=f_a e^{ia(x)/f_a} . 
\end{equation}
 When the PQ transition occurs, the potential energy with  different values of  $a$ is nearly degenerate, so $a$ is expected to take on a   random initial value. The expansion of the universe will smooth out spatial variations in $a(x)$ but the average value of $a(x)$ remains random until late times. We say the field is misaligned with respect to its minimum, and the  energy stored in this misalignment will eventually become the dark matter.  There are two    different cases for the cosmological evolution. In  case 1, the  reheating temperature of inflation is less than $T_{\text{PQ}}$ and the PQ symmetry is broken during inflation and never restored afterwards. In this case the axion field is smoothed during inflation and randomly obtain a spatially uniform vacuum expectation value $\alpha f_a$, where $\alpha$ is known as the misalignment angle. Quantum fluctuations in $a$ are small and proportional to the Hubble scale during inflation. As these fluctuations in $a(x)$ are isocurvature, and the cosmic microwave background observations place a strong limit on   isocurvature fluctations, in case 1 there is a strong upper bound on the scale of inflation\cite{Linde:1985yf,Seckel:1985tj,Lyth:1989pb,Turner:1990uz,Lyth:1992tx,Fox:2004kb}.  In case 2 the reheating temperature after  inflation is greater than $T_{\text{PQ}}$,  and the PQ symmetry breaks after inflation. In this case the axion takes on random values uncorrelated over scales which are larger than the Hubble horizon at the time of PQ breaking. Topological   axion strings and domain walls are then formed after inflation. Provided that there is no nontrivial unbroken discrete subgroup of the PQ symmetry, every domain wall ends on an axion string and the whole network of strings and domain walls will eventually disappear\cite{Sikivie:1982qv,Chang:1998tb,Gorghetto:2018myk}.   The cosmological  restriction that in case 2 the PQ symmetry must not have any exact discrete subgroup is a severe but achievable constraint on axion model building.

The evolution equation of the axion field in the early universe can be described by the equation
\begin{equation}\label{eq1}
\left(\partial_t^2+3\frac{\dot{R}}{R}\partial_t-\frac{1}{R^2}\nabla_x^2\right)a(x)+V'(a)=0
\end{equation}
where $R$ is the scale factor, the components of $\textbf{x}$ are the co-moving spatial coordinates of the universe, and $V(a)$ is the effective potential energy density of the axion field. This potential comes from non-perturbative QCD effects such as instantons\cite{tHooft:1976snw}, which break the $U_{PQ}(1)$ symmetry to a $Z(N)$ discrete subgroup\cite{Sikivie:1982qv}. In case 2, we must have $N=1$ in order to avoid overclosure of the universe by a frustrated network of axion strings and domain walls, while in case 1 any such defects are inflated away (however, see ref. \cite{Kaplan:2008ss} for a conceivably observable effect of axion strings outside our horizon). We can write the instanton potential qualitatively as: 
\begin{equation}
V_a=f_a^2m_a^2(T)\left[1-\text{cos}\left(\frac{a}{f_a}\right)\right]
\end{equation}
where $m_a$ is the axion mass, which is a function of temperature $T$. The cosine form  comes from the dilute instanton gas approximation and is not exact. The form of the axion potential at low temperatures may be found in reference \cite{Braaten:2016kzc}.
At high temperature ($T>$1 GeV), $m_a(T)$ can be estimated by instanton effects and by lattice QCD.  While there is disagreement between different approaches these disagreements will not significantly change our results  \cite{Dine:2017swf}.  
The axion mass is   constant when $T$ is below the QCD scale and the calculation  at low energies  is reliable due to chiral perturbation theory. However, we cannot reliably predict the axion mass  when $T$ is between 0.2 GeV and 1 GeV.  In standard thermal history, this uncertainty  will not affect our prediction of the axion relic density because the temperature at the critical time is higher than 1 GeV. However, we will see in the next section that   early matter domination will decrease the critical temperature.  We will assume the axion mass to be a continuous function of $T$,  whose exact form will not change our main results. The full expression we will use  for the  axion mass follows ref. \cite{Hertzberg:2008wr}:
\begin{equation}
m_a(T)=
\begin{cases}
m_a(0), \quad\quad\quad\quad\quad\quad\quad\quad T<0.2\ \text{GeV} \\
 m_a(0)(\frac{0.2\ \text{GeV}}{T})^{6.5}, \quad 0.2\ \text{GeV}\leq T\leq1\ \text{GeV}\\
b m_a(0)(\frac{0.2\ \text{GeV}}{T})^4, \quad\quad\quad\quad T>1\ \text{GeV}\ ,
\end{cases}
\end{equation}
where $b=0.018$ and $m_a(0)=(78\ \text{MeV})^2/f_a$, and $m_a(0)$ is the axion mass at zero-temperature. Given the thermal history of early universe, the axion mass is determined by cosmic time. The first three terms in Eq.(\ref{eq1}) are proportional to $t^{-2}$, which are the dominant terms until late times. We   define the critical time $t_{1}$ at which the potential term becomes important relative to the Hubble expansion term to be:
\begin{equation}
H(t_1)=m_a(T(t_1))
\end{equation}
The mean value of the axion field does not evolve much before the time $t_1$. After $t_1$ the axion field begins to oscillate and its energy density behaves approximately like nonrelativistic   matter. The energy density of a uniform oscillating axion field may  be interpreted as the energy density of axion particles at rest.
The number of  axion particles per co-moving volume is  adiabatically conserved  because the axion mass changes slowly compared with the oscillation period. In case 1, where axions were homogenized by inflation, axions at rest  are the dominate initial component of axions in the universe. In case 2, some spatial variation in the axion field remains which is interpreted as axions with non zero momentum, and also a substantial number of axions are produced via the decay of axion strings and domain walls. The number density of  axions at rest is\cite{Preskill:1982cy,Abbott:1982af,Dine:1982ah}:
\begin{equation}
n_a^{vac,0}(t)=\frac{1}{2}m_a(t) f_a^2 \alpha^2\ .
\end{equation}
 In case 1,    $\alpha$ is uniform throughout our universe and its random value introduces uncertainties in our prediction. We simply treat it as a $O(1)$ constant and do not consider the possible consequences of a  small misalignment angle. In case 2, $\alpha$ is randomly distributed taking on many different values   throughout our observable universe,  and is roughly uniform on scales on the order of the Hubble horizon size at the time of PQ symmetry breaking. As there are many such volumes contained within our current horizon we may average over the different initial values. The dominant source of theory uncertainty for the axion density in case 2 is from the  computation of the number of axions produced from  the decay of axion strings and domain walls.

\subsection{Axion Relic Density}

In case 1, we can directly get the current energy density of the axion:
\begin{equation}
\rho_a^{\text{vac},0}=\frac{1}{2}m_a(0)m_a(t_1)f_a^2\alpha^2\left(\frac{R_1}{R}\right)^3
\end{equation}
where $t_1$ is the critical time when axion starts to oscillate and $R_1/R$ is the ratio of the scale factor at the critical time to that at present. The number of axions is  approximately conserved and the energy density is simply the number density multiplied by $m_a(0)$. Combined with a  radiation dominated thermal history, we obtain the following energy density in case 1:
\begin{equation}\label{eq3}
\Omega_a\sim0.15\:\left(\frac{f_a}{10^{12} \text{GeV}}\right)^{7/6}\left(\frac{0.7}{h}\right)^2\alpha^2
\end{equation}
where $h$ is defined to give the Hubble constant $H_0=\text{100km/s}\cdot h\cdot\text{Mpc}$.

Case 2 is more complicated because axion strings and  domain walls will decay to axions and give  an extra contribution to the axion relic density. There is a potential so-called domain wall problem when the  PQ symmetry group $U_{PQ}(1)$ has non trivial discrete $Z(N)$ subgroup, as in this case there is an N fold degeneracy of the vacuum\cite{Sikivie:1982qv}. We assume in case 2 that $N=1$ for viable axion cosmology \cite{Vilenkin:1982ks,Lazarides:1982tw}. In this case  the domain walls are unstable and bounded by strings. The string decay contribution to the axion relic density \cite{Gorghetto:2018myk} is highly uncertain and we simply parameterize the uncertainty. Following  ref. \cite{Hagmann:1990mj} we write:
\begin{equation}
\rho_a^{\text{str}}=m_a(0)n_a^{\text{str}}(t_1)\left(\frac{R_1}{R_0}\right)^3\backsimeq Y m_a(0) \frac{f_a^2}{m_a(t_1)}\left(\frac{R_1}{R_0}\right)^3
\end{equation}
where $m_a(0)$ is the axion mass at zero temperature, $m_a(t_1)$ is the axion mass at the critical time when axion starts to oscillate, and $Y$ is an order one factor which is determined by   details  such as the efficiency of string decay, the axion string number per horizon and average energy of the axions emitted in a string decay.   We simply assume a value for $Y$ here and study what will be different in a nonstandard thermal history. 

The last step is to estimate the contribution from higher momentum modes. Assume that axion field varies by $f_a$ from one horizon to the next, we can obtain the number density distribution of  higher momentum  modes of the axion:
\begin{equation}\label{eq2}
\frac{n_a}{d\omega}\sim \frac{f_a^2}{2t^2\omega^2}
\end{equation}
Only frequencies which enter the horizon are physically relevant for this work. Integrating over $\omega>1/H(t_1)$ in Eq.(\ref{eq2}) gives us the contribution from vacuum realignment of higher momentum modes:
\begin{equation}
\rho_a^{\text{vac},1}\sim \frac{m_a(0) f_a^2}{2m_a(t_1)}\left(\frac{R_1}{R_0}\right)^3
\end{equation}
So the contribution from higher momentum modes is roughly the same as that of the zero momentum mode.
Including an estimate of the contribution from higher momentum modes and string decays, the relic density could be written as:
\begin{equation}\label{eq4}
\Omega_a\sim0.6\:\left(\frac{f_a}{10^{12} \text{GeV}}\right)^{7/6}\left(\frac{0.7}{h}\right)^2
\end{equation}
Notice that the axion relic density in case 2 is generally greater than  in case 1 for a given $f_a$.

From Eqs.(\ref{eq3}),(\ref{eq4}), we see an upper bound for $f_a$ in order to avoid overproduction of axion dark matter. The upper bounds for case 1 and case 2 with standard thermal history are respectively $\sim1.4\times10^{12}$ GeV and $\sim4.4\times10^{11}$ GeV, with order one uncertainties in both cases.  Combined with other constraints, we obtain the so-called axion window, $10^9\ \text{GeV}<f_a<10^{12}\ \text{GeV}$. 

\subsection{Axion Miniclusters}
In case 2, where inflation happens before the PQ phase transition, the initial misalignment angle  will not be homogenized by inflation. Therefore, its value will vary randomly from one horizon to another.  An inhomogeneity with $\delta \rho_a/\rho_a= \mathcal{O}(1)$ is produced when the axion mass   turns on. If not erased by the free-streaming effect, gravitationally bound objects , which are called axion miniclusters, may form at the time $t_{eq}$ when energy density of radiation and matter are equal\cite{Hogan:1988mp,Kolb:1993zz,Kolb:1995bu,Chang:1998tb}.

Because axions are typically very cold,  free-streaming effects will not restrain the form of axion miniclusters\cite{Kolb:1995bu,Chang:1998tb}. In case 1, only zero mode axions due to   vacuum misalignment are produced and there is no velocity dispersion. In case 2, there are higher momentum modes produced by vacuum realignment axions produced by wall decay and string decay. They will give us some non-zero velocity dispersion but it can still be shown that free-streaming will not homogenize the axions.

The characteristic minicluster mass is given by the total mass of axions contained within the horizon at the critical time when the axion starts to oscillate:
\begin{equation}\label{Mmc}
M_{mc}=\frac{1}{2}m_a(0)m_a(t_1)f_a^2\:\frac{4\pi}{3}\left(\frac{1}{H(t_1)}\right)^3
\end{equation}
Since the number of axions per co-moving volume is conserved, we must take the evolution of the axion mass into consideration in computing the mass of axion miniclusters.   
If we assume a standard thermal history where the early universe is dominated by radiation, the corresponding temperature of the critical time is:
\begin{equation}
T_1\simeq 1\:\text{GeV}\left(\frac{10^{12}\text{GeV}}{f_a}\right)^{1/6}
\end{equation}

Thus we obtain the mass of axion miniclusters in a standard thermal history:
\begin{equation}\label{mass}
M_{mc}=3.7\times10^{-10}M_{\odot}\left(\frac{f_a}{10^{12}\text{GeV}}\right)^{5/3}
\end{equation}
where $M_{\odot}$ is the solar mass. Note that there are various strategies to estimate the mass of axion miniclusters, such as calculating the axions contained within the horizon at $t_{eq}$. A detailed study requires the calculation of the mass function of axion minicluster through its power spectrum. The evolution of axion miniclusters in the nonlinear region should be also included, which allows for the possibility of larger axion stars. Such evolution is outside the scope of this paper. Our main goal is to find what will be changed during a nonstandard thermal history. Therefore, we focus on the linear region and give the estimate of the change in the initial axion minicluster mass   with early matter domination.

\section{Opening  the Axion Window}\label{s2}
Nothing so far has been   directly  detected from the epoch after  inflation and before nucleosynthesis. The ``standard''  assumption  about that period is that the inflationary energy density  decayed to  a  hot thermal relativistic plasma containing all the particles in the standard model and possibly some extension\cite{Albrecht:1982mp,Turner:1983he,Traschen:1990sw,Kofman:1994rk}, reheating the universe,  and the universe remained radiation dominated until the temperature dropped below about 1 eV. However,  the inflationary energy density could also decay to some nonstandard massive particles, which could be long lived and come to dominate the energy of the universe,  as the energy density of nonrelativistic particles evolves with the scale factor $R$ as $R^{-3}$ while that of radiation evolves as $R^{-4}$. The success of standard nucleosynthesis implies that any such  massive long lived particles   have  decayed and brought the universe to radiation domination  before a temperature of order a few MeV. A  pre-nucleosynthesis epoch with energy  dominated by nonrelativistic massive particles is called  an Early Matter Dominated Epoch (EMDE). Such a scenario is favored by some   baryogenesis  models as a way to satisfy the out of equilibrium  Sakharov condition  for producing the  asymmetry between matter and anti-matter.  One top down motivated example of an EDME is motivated by stabilized moduli in string theory \cite{deCarlos:1993wie,Banks:1993en,Acharya:2010af}. Another motivation is to produce curvature perturbations in the curvaton model\cite{Mollerach:1989hu,Linde:1996gt,Lyth:2001nq,Moroi:2001ct}. There are  other cosmological consequences of a EDME scenario, such as boosting the thermal dark matter annihilation rate\cite{Khlopov:1982ef,Erickcek:2015bda,Erickcek:2015jza,Choi:2017ncz}.

An EDME will affect the axion relic density and expand the allowed range of $f_a$.  The co-moving entropy density  will increase during the decay of the massive particles, which will decrease the ratio of axions to photons  \cite{Giudice:2000ex,Grin:2007yg,Visinelli:2009kt,Kane:2015jia,Lazarides:1990xp}.
We may calculate  the abundance of relic axions with an EDME by solving  Boltzmann equations. We will use the example of matter domination by a   particle $\Phi$, whose spin  is irrelevant. In the three-fluid model for reheating, the evolution of energy density is give by:
\begin{equation}
\begin{split}
&\frac{d\rho_{\Phi}}{dt}+3H\rho_{\Phi}=-\Gamma_{\phi}\rho_{\Phi}\\
&\frac{d\rho_{r}}{dt}+4H\rho_{\Phi}=\Gamma_{\phi}\rho_{\Phi}
\end{split}
\end{equation}
where $\rho_{\Phi}$ is the energy density of  $\Phi$, $\rho_r$ is the energy density of radiation, $\Gamma_{\Phi}$ is the decay rate of the massive particle and $H$ is the Hubble parameter. We have neglected the contribution from the axion field because it contributes only a minor energy density to  the early universe. Combined with the Friedmann equations we can solve the exact energy density of the massive particle and radiation as a function of cosmic time. We assume that the radiation plasma reaches its equilibrium state instantaneously after $\Phi$ decays. This is reasonable since the decay rate is relatively slow compared with the thermalization rate of the light particles. In this way we can also obtain the temperature in terms of cosmic time, which also gives us the mass of the axion as a function of time. Once we know the axion mass as a function of time, the critical time $t_1$ when the axion starts to oscillate can be estimated from $H(t_1)=m_a(t_1)$. We then use the adiabatic  approximation with the co-moving number density of axions  conserved after the critical time, which gives the evolution of axion number density at time $t>t_1$:
\begin{equation}
n_a(t)=\frac{1}{2}m_a(t_1)f_a^2\left(\frac{R(t_1)}{R(t)}\right)^3\ .
\end{equation}
 When the universe is dominated again by radiation, the entropy density behaves exactly like $R^{-3}$ and $n_a/s$ is conserved. The entropy of universe is dominated by radiation. We can thus obtain the current  axion density.  The axion energy density must be less than the dark matter relic density. We can therefore obtain an upper bound on the axion decay constant $f_a$  as a function of the reheat temperature $T_{rh}$. The reheat temperature is directly determined by the decay rate of oscillating scalar field.
\begin{equation}
\frac{\pi^2}{30}g_{*}(T_{rh})T_{rh}^4=\frac{3E_{pl}^2\Gamma_{\phi}^2}{8}
\end{equation}
where $g_{*}$ is the effective number of degrees of freedom, and $E_{pl}$ is the planck energy.
\begin{figure}
  \centering
  \includegraphics[width=9cm]{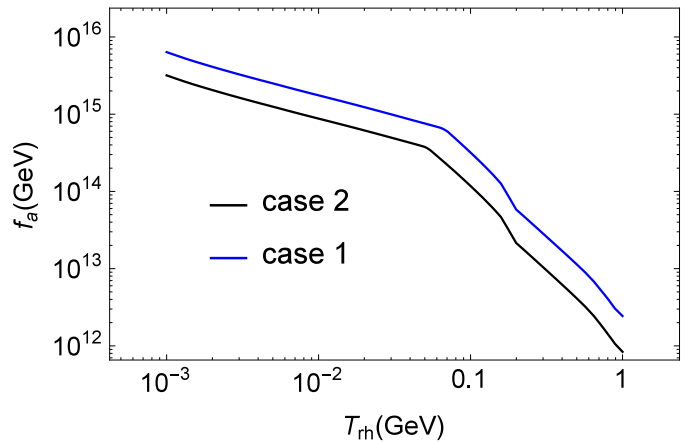}\\
  \caption{New upper bound on the axion decay constant under early matter domination with reheating temperature $T_{rh}$, assuming the various undetermined constants are of order 1. The blue curve is for case 1 and black curve is for case 2. }\label{1}
\end{figure}
The upper bound on $f_a$ does not depend on the EDME   unless the reheat  temperature is below   the temperature at  the critical time when the axion field begins to oscillate. Thus, only  a reheat temperature greater than about 1 MeV (So it happens before nucleosynthesis) and less than about 1 GeV is relevant for a new story of axion cosmology.

\section{Axion Miniclusters with Early Matter Domination}\label{s3}
Since the axion window is widened by early matter domination, it is straightforward to show that the possible mass of axion miniclusters increases with a greater axion decay constant. A  nontrivial result is that the formation of axion miniclusters is even allowed in case 1, which is not expected  with a standard thermal history. The formation of miniclusters results because matter density perturbations will grow linearly with the scale factor during the EMDE while they only  grow  logarithmically during radiation domination. Generally   an EDME allows for an increase in small scale dark matter structure formation\cite{Erickcek:2011us,Fan:2014zua}. But axion miniclusters are especially sensitive to such a period because axions are extremely cold.  Free streaming effects are negligible for axions which allows for tiny structures.

For miniclusters in case 2, the correction  to the axion minicluster mass from early matter domination is straightforward. We simply estimate the critical time with early matter domination and use the same formula Eq.(\ref{Mmc}). However, in case 1 with standard thermal history axion miniclusters do not generally form. In contrast, the primordial perturbations of axion field that enter the horizon  during the EDME can grow linearly, and form an axion minicluster. If such structures form during the EDME then  they are  dominated by the massive particles which will decay to reheat the universe, and when those particles decay   the structures will be erased. But structures that form during and after  the radiation dominated epoch  persist.  We estimate the initial size of such structures as follows.  We assume a nearly scale invariant  primordial perturbation of   about $10^{-4}$\cite{Komatsu:2010fb}. Axions are frozen before the critical time at which axion oscillations begin.  When there is a period of EDME and  the  reheating time is later than  the critical time,   initial  inhomogeneities which are inside the horizon will grow linearly with the scale factor.  We therefore find the scale at which perturbations of a given scale size enter the horizon during the EDME, and grow to $\delta\rho/\rho$  of order 1  at the end of the early matter domination epoch. Continued logarithmic growth of these structures will allow for   axion minicluster  formation at the end of radiation domination. 
Only  specific combinations of $f_a$ and $T_{rh}$ will allow the formation of axion miniclusters in case 1. In general  larger axion decay constants lead to a   later   critical time at which the axion starts to oscillate, and these structures grow linearly with the scale factor only during  the EDME. In case 1,   formation of axion miniclusters implies a reheating temperature dependent upper bound on $f_a$.   

\subsection{Cosmological Perturbations}
In order to obtain the axion minicluster mass in case 1, we need to find the perturbation growth during early matter domination. Generally the perturbations will grow linearly with the scale factor if the universe is dominated by matter. In principle the situation  is more complicated for the axion because its mass is also changing with time, however the term from the changing mass is negligible compared with the linear growth term for the following reasons: 1. The temperature is typically less than 1 GeV at the critical time when the axion perturbation starts to grow. The axion mass will not change much at that time. 2. The axion mass is temperature dependent and gives perturbations proportional to $\dot{T}/T$, which is actually a logarithmic growth term.
We can treat the oscillating scalar field, the radiation and cold axions as perfect fluids with energy momentum tensors\cite{Kodama:1985bj,Malik:2002jb,Lemoine:2006sc}:
\begin{equation}
T^{\mu\nu}=(\rho+p)u^{\mu}u^{\nu}+pg^{\mu\nu}
\end{equation}
Where $u^{\mu}\equiv dx^{\mu}/d\lambda$ is the four-velocity. For cold axions and $\Phi$ particles, the pressure is zero and for radiation $p=\rho/3$. Due to the decay of $\Phi$ particles, different fluids exchange energy covariantly:
\begin{equation}
\nabla(^{(i)}T^{\mu}_{ \nu})=Q^{(i)}_{\nu}
\end{equation}
Where $i$ denotes different fluids. For the energy exchange vector:
\begin{equation}
\begin{split}
&Q^{\phi}_{\nu}={}^{(\phi)}T_{\mu\nu}u^{\mu}_{\phi} \Gamma_{\phi}\\
&Q^{r}_{\nu}=-Q^{\phi}_{\nu}-Q^{a}_{\nu}\\
&Q^{a}_{\nu}=-{}^{(a)}T_{\mu\nu}u^{\mu}_{a}\frac{\dot{m}_a}{m_a}
\end{split}
\end{equation}
During the early matter domination, $Q^{a}\ll Q^{\phi}$. So the the perturbation  in axions should not change the evolution of the radiation perturbation. To obtain the perturbation equations, we start with the perturbed metric
\begin{equation}
\text{d}s^2=-(1+2\Psi)\text{d}t^2+a^2(t)\delta_{ij}(1-2\Psi)dx^{i}dx^{j}
\end{equation}
Thus we have the perturbation of the four-velocity:
\begin{equation}
\begin{split}
&u^0=1-\Psi\\
&u^j_{i}=(1-\Psi)V_{(i)}^j
\end{split}
\end{equation}
where $V_{(i)}^j\equiv \text{d}x^j/\text{d}t$ is the fluid velocity of the $i$th fluid. With the perturbation of energy density of each fluid $\rho_i=\rho^0_i(1+\delta_i)$, we can write the dominant term and the first order perturbation term of $Q$
\begin{equation}
\begin{split}
&Q^{(\phi)}_0=\Gamma_{\phi}\rho^0_{\phi}(1+\delta_{\phi}+\Psi)\\
&Q^{(\phi)}_j=-\Gamma_{\phi}\rho^0_{\phi}a^2\delta_{kj}V^k_{\phi}\\
&Q^{(a)}_0=-\frac{\dot{m}_a}{m_a}\rho^0_{a}(1+\delta_{a}+\Psi)\\
&Q^{(a)}_j=\frac{\dot{m}_a}{m_a}\rho^0_{a}a^2\delta_{kj}V^k_{a}
\end{split}
\end{equation}
$\Gamma_{\phi}$ and $\frac{\dot{m}_a}{m_a}$ are significantly different. One is a constant and the other has perturbation determined by the temperature perturbation of the radiation. Compared with the Hubble parameter, $\Gamma_{\phi}$ is usually negligible but $\frac{\dot{m}_a}{m_a}$ may be important for the perturbation function.

Expressing $Q_{\nu}$ with  in terms of the zero-order and first-order perturbations, we can combine equation (1) and (2) to get simple results that determine the perturbation:
\begin{equation}
\begin{split}
&\frac{\text{d}{\delta}}{\text{d}t}+(1+w)\frac{\theta}{a}+3(1+w)\frac{\text{d}\Psi}{\text{d}t}=\frac{1}{\rho^0}[Q_0^{(0)}\delta-Q_0^{(1)}]\ , \\
&\frac{\text{d}\theta}{\text{d}t}+(1-3w)H\theta+\frac{\nabla^2\Psi}{a}+\frac{w}{1+w}\frac{\nabla^2\delta}{a} \\ &=\frac{1}{\rho_0}\left[\frac{\partial_i Q_i}{a(1+w)}+Q_0^{(0)}\theta\right]\ ,
\end{split}
\end{equation}
where $w=p/\rho$ is the fluid's equation of state parameter, and  $\theta=a\partial_i V^i$ is the divergence of fluid's conformal velocity. $Q_0^{(0)}$ and $Q_0^{(1)}$ are respectively the zero-order and first-order components of $Q$.

It can be generally shown that the metric perturbation  is frozen in a matter-dominated universe.
We define the beginning of early matter domination as $a=1$ and its corresponding Hubble parameter is $H_0$. For convenience, we also define dimensionless parameter $\widetilde{\theta_{\phi}}\equiv \theta_{\phi}/H_0$, $\widetilde{k}\equiv k/H_0$.
Therefore we can represent our equations during early matter domination in the following way:
\begin{equation}
\begin{split}
&a^{-1/2}\delta^{'}_{\phi}(a)+\widetilde{\theta}_{\phi}(a)=0\ ,\\
&a^{1/2}\widetilde{\theta}_{\phi}^{'}(a)+a^{-1/2}\widetilde{\theta}_{\phi}(a)+\widetilde{k}^2\Psi=0\ ,\\
&a^{1/2}\widetilde{\theta}^{'}_a(a)+a^{-1/2}\widetilde{\theta}_{a}(a)+\widetilde{k}^2\Psi=0\ ,\\
&a^{-1/2}\delta^{'}_{a}(a)+\widetilde{\theta}_{a}(a)=\frac{a\dot{m}_a}{m_a H(t_1)}\Psi\ , \\
\end{split}
\end{equation}
where a prime represents the derivative to scale factor, and $H(t_1)$ is the Hubble parameter at the critical time. It is not hard to show that the term $\frac{a\dot{m}_a}{m_a H(t_1)}$ only causes a logarithmic growth, which could be neglected compared with the linear growth. Eventually the perturbation  for modes that have already entered horizon before the critical time is:
\begin{equation}
\delta_a(a,k)=2\Psi_0+\frac{2k^2}{3H(t_1)^2}a\Psi_0
\end{equation}
where $\Psi_0$ represents the primordial perturbation of quantum fluctuation during inflation, which is about $10^{-4}$. Now it is clear that perturbations grow linearly with scale factor $a$ during early matter dominated epoch. To form axion miniclusters efficiently at the end of radiation domination, $\delta_a$ must be grow up to about 1. Actually the formation of axion minicluster is complicated here because the growth depends on momentum. We can actually obtain the transfer function for axion generally and calculate the mass function of axion minicluster with Press-Schechter formalism. However, axion perturbation grows to the nonlinear region   very early  and it is hard to predict its later evolution. In this paper we just estimate the original axion minicluster mass and leave its evolution for future research.

\subsection{Formation of Axion Miniclusters}
\begin{figure}
  \centering
  \includegraphics[width=9cm]{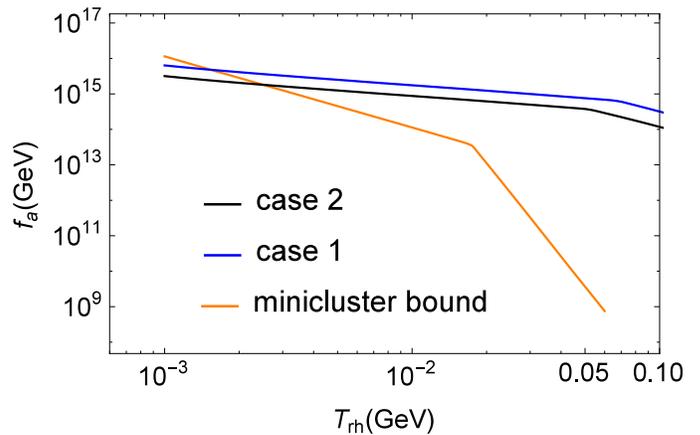}\\
  \caption{The blue and black curve which indicate the bound on $f_a$ from the relic axion density is part of FIG.(\ref{1}). The orange curve represents the upper bound for axion decay constant if growth is sufficient for formation of axion miniclusters in case 1. (In case 2, due to larger initial inhomogeneity on small scales, axion miniclusters can generally form.) The   axion density increases with the axion decay constant. When the reheating temperature is sufficiently low,   dark matter axions can comprise all of the dark matter as well as form axion miniclusters  in case 1,  because the blue   curve is below the orange curve.}\label{2}
\end{figure}

A perturbation $\delta_a(a,k)$ will  start to grow when $k$ enters the horizon. The momentum modes which have already entered the horizon at the critical time ($k>H(t_1)$)  will grow the largest. Typically large $k$ represents smaller axion miniclusters so we only care about $k< H(t_1)$. Therefore the criterion for axion miniclusters formation in case 1 is if $\delta_a(a_e,H(t_1))$ is larger than 1, where $a_e$ is the scale factor at the end of early matter domination.  It can be drawn together with the upper bound for allowed axion density (See FIG.(\ref{2})). Parameters must be below the bound from axion density to not overproduce axions. To form axion miniclusters, the parameters must be below the orange curve. For case 1, if all the dark matter is axions, the reheating temperature  must be below about 60 MeV in order to obtain axion miniclusters.

\begin{figure}
  \centering
  \includegraphics[width=9cm]{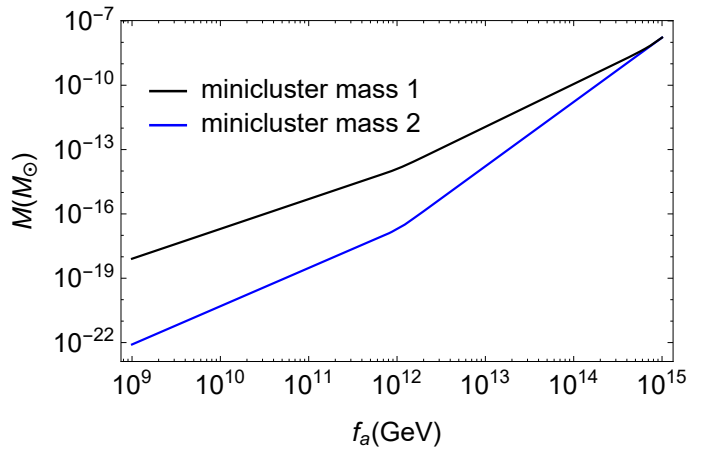}\\
  \caption{The black curve and the blue curve respectively represent the axion minicluster mass in case 1 and case 2 when the reheating temperature is 3 MeV. }\label{3}
\end{figure}

The final step of this chapter is to determine the mass of axion miniclusters with early matter domination. In case 2 it is can be straightforwardly done by substituting the new critical time. In case 1, suppose that we have some $k_c< H(t_1)$ which satisfies:
\begin{equation}
\delta_a(a_e,k_c)=1
\end{equation}
where $a_e$ is the scale factor at the end of early matter domination. $k_c$ represents the characteristic modes that eventually grow to axion miniclusters. Suppose that $k_c$ enters the horizon at time $t_c$, $H(t_c)=k_c$.
The corresponding axion number density at $t_c$ is:
\begin{equation}
n_{ac}=\frac{1}{2}m_a(t_1)f_a^2(\frac{R(t_1)}{R(t_c)})^3
\end{equation}
where $t_1$ is the critical time and $R$ is the scale factor of universe. Therefore axion minicluster mass mass could be estimated by:
\begin{equation}
M_{mc}=\frac{4}{3}\pi m_a(0)(\frac{1}{k_c})^3 n_{ac}
\end{equation}
As an example, we calculate how axion minicluster mass changes with axion decay constant at reheating temperature $3$ MeV, as shown in FIG.(\ref{3}).
From FIG.(\ref{3}) we can see that the upper limit on the axion minicluster mass has increased to $10^{-8} M_{\odot}$, where $M_{\odot}$ is the solar mass.  In comparison, from Eq.(\ref{mass}), with the standard thermal history  the maximum minicluster mass is $\sim 8.5\times10^{-14} M_{\odot}$. It is worth noting that the initial axion minicluster mass is typically less than the critical mass  at which an axion star becomes unstable to gravitational collapse\cite{Visinelli:2017ooc,Helfer:2016ljl,Chavanis:2016dab,Michel:2018nzt} so the miniclusters will not form black holes. 
 \section{Conclusions}
We have shown that the axion window is wider and the formation history of axion miniclusters is significantly affected by a period of matter domination prior to nucleosynthesis. The axion  can be lighter, and the maximum  mass of axion miniclusters is increased. Furthermore axion miniclusters can form even in the case where the PQ symmetry breaking occurs before inflation.   In this work we have estimated the characteristic mass of axion miniclusters at the time of formation. More detailed numerical study about the evolution of axion miniclusters is needed to obtain the information about axion miniclusters at present, including the mass function of axion miniclusters, the percentage of axions that form axion miniclusters, and the percentage of axion miniclusters that form boson stars or black holes. The evolution of axion miniclusters after their formation and the fraction of axions that finally become gravitationally bound objects requires detailed numerical study, which is  beyond the scope of this paper. Such work  would be important, as indirect detection of axion miniclusters could possibly provide evidence for both the existence of the axion and for  a nonstandard thermal history of the very early universe.
\section{Acknowledgements}
This work was supported in part by the by the DOE under grant  DE-SC0011637 and by the Kenneth K. Young Memorial Endowed Chair.

\bibliographystyle{apsrev}
\bibliography{axionminiref}

\end{document}